\documentclass[prb,twocolumn,showpacs,amsmath,amssymb]{revtex4-1}
\usepackage{fullpage}
\usepackage{graphicx}
\usepackage{epstopdf}
\usepackage{dcolumn}
\usepackage{multirow}
\usepackage{color,amsmath,amssymb}
\usepackage{bm}
\begin{document}

\title{Exchange  constants in molecule-based magnets  derived from
  density functional methods}
\author{I.O. Thomas}
\author{S.J. Clark}
\author{T. Lancaster}
\affiliation{Durham University, Centre for Material Physics,
  Department of Physics, South Road, Durham, DH1 3LE, UK}
\date{\today}

\begin{abstract}  
Cu(pyz)(NO$_{3}$)$_{2}$ is a quasi one-dimensional molecular
antiferromagnet that exhibits three dimensional long-range magnetic
order below $T_{\mathrm{N}}=110$~mK due to the presence of weak
inter-chain exchange couplings. Here we compare calculations of the
three largest exchange coupling constants in this system using two
techniques based on plane-wave basis-set density functional theory:
(i) a dimer fragment approach and (ii) an approach using periodic
boundary conditions.  
The calculated values of the large intrachain coupling constant are
found to be consistent with experiment, showing the expected level of variation
between different techniques and implementations. 
However, the interchain coupling constants are found to be smaller than
the current limits on the resolution of the calculations.
This is
due to the computational
limitations on convergence of absolute energy differences with respect
to basis set, which are larger than the inter-chain couplings
themselves. Our results imply that errors resulting from such
limitations are inherent in the evaluation of small
exchange constants in systems of this sort, and that many previously
reported results should therefore be treated with caution.
\end{abstract}
\maketitle

\section{Introduction}
Molecule-based magnets \cite{Blundell-2004} provide experimental
realisations of magnetic spin systems that were, until recently, the
sole preserve of theorists \cite{Lancaster-2013}.  They allow an
opportunity to investigate magnetism in low-dimensional systems, the
occurrence of quantum critical points, the possible existence of spin
liquid states, along with many other phenomena \cite{Lancaster-2013}.
The theoretical characterisation and classification of these materials
in terms of their dimensionality, the phases they display and their
low-energy magnetic behaviour is an important part of this
investigation.  At low energies, the properties of a magnetic material
may be mapped onto those of a magnetic spin model
\cite{Whangbo-2003,Nova-2011,Moreira-2006,Datta-2015}, (e.g.\ the
Heisenberg model) in which the chemical, structural and electronic
properties underlying its magnetism are encoded in a small number of
parameters, such as exchange constants.  The diversity of possible
structural and chemical arrangements allows for the synthesis of
molecule-based compounds whose behaviour in this energetic limit
corresponds to many different model systems.  Matching a given
material with the most appropriate magnetic model is therefore
important if its properties are to be properly explored and
understood.

{\em Ab initio} techniques such as density functional theory (DFT)
\cite{Martin-2004,Lejaeghereaad-2016}, and semi-empirical methods
based on DFT, such as some hybrid functionals or DFT+U, are often used
to determine the appropriate low-energy model for a given
material. This involves determining the relative energies of spin
configurations, which are used to extract coupling constants.  Two
frequently employed methods of doing this are the dimer fragment
approach (DFA) \cite{Whangbo-2003,Nova-2011} and the periodic approach
(PA)\cite{Whangbo-2003, Moreira-2006}.

In the DFA it is assumed that, for a given exchange pathway between
magnetic centres at sites $a$ and $b$, the most significant
contributions to the exchange constant will be from {\em only} those
sites and their associated ligands (that is, the exchange constant is
dependent only on properties local to $a$ and $b$).  Calculations (see
e.g. Ref.~\onlinecite{Moreira-2006} and references therein) show that
this can be a reasonable assumption, but for magnetic structures whose
relative energies are small (leading, for example to exchange
constants of the order of $\lesssim 0.1$ meV) it is likely that
systematic errors that are introduced by focusing on a dimer fragment
of the system, rather than the system as a whole, will be significant.
In contrast to the DFA, the PA is the method of simulating crystalline
systems using periodic boundary conditions.  This is standard
technology in plane-wave basis set electronic structure codes, where
crystalline systems are simulated (see, for example,
Ref.~\onlinecite{Payne-1992}, section IIC). Interactions between all
of the electrons and nuclei of the bulk system are taken into account.
More generally, for both DFA and PA, we might also ask whether we face
cases where the energetic separation of different magnetic states is
smaller than the energy resolution of a DFT implementation.  This can
lead to situations, for example, where different calculations disagree
on whether a small exchange constant is ferromagnetic or
antiferromagnetic, which would be reflected in very different
predictions of the resulting magnetic behaviour. 

In this paper, we compare properties computed using the DFA with those
computed using the PA to assess the effect of significant systematic
errors, whose neglect we suspect to be widespread and which are rarely
addressed in the literature. The main purpose of this paper is to
examine the numerical errors that are incurred in electronic structure
calculations rather than the systematics of various Hamiltonians
commonly used in DFT, such as (semi-)local functionals (for example
PBE), non-local hybrid functionals (for example B3LYP, HSE) or Hubbard
contributions (for example DFT+U). The application and performance of
various functionals have been examined
elsewhere\cite{Moreira-2006,Jornet-Somoza-2010,DosSantos-2016,feng-2004}.

We have selected the quasi one-dimensional (1D) Heisenberg
antiferromagnet Cu(pyz)(NO$_3$)$_2$ (copper pyrazine
dinitrate)\cite{Santoro-1970} as the basis of a case study, and
perform this comparison for its three largest exchange constants.
Cu(pyz)(NO$_3$)$_2$ has been chosen because it provides a good
experimental realisation of the $S=1/2$ 1D antiferromagnetic
Heisenberg model \cite{Villa-1971,Losee-1973}.  As a result, its
magnetic \cite{Villa-1971,Losee-1973,Mennenga-1984,
  Hammar-1999,Lancaster-2006,Jornet-Somoza-2010}, spin-dynamic
\cite{Stone-2003,Kuhne-2009,Kuhne-2010,Kuhne-2011,Rohrkamp-2010,
  Gunyadin-Sen-2010}, thermal \cite{Sologubenko-2007,Shimshoni-2009}
and vibrational \cite{Jones-2001,Brown-2007} properties have been the
subject of experimental and theoretical investigation.

Cu(pyz)(NO$_3$)$_2$ in a coordination polymer consisting of a crystalline arrangement of chains of
Cu$^{2+}$ ions linked by pyrazine (pyz) ligands parallel to the $a$
axis of its unit cell, where each Cu$^{2+}$ ion is also bonded to a
pair of nitrate ions, illustrated in Fig.~\ref{fig:basicCell}.  The
dominant magnetic exchange interaction, with exchange constant $J_{\rm
  d2}$ (see below), results from superexchange between the adjacent
Cu$^{2+}$ ions in each chain, mediated by the pyrazine ligands
\cite{Richardson-1976}.  Inter-chain magnetic interactions, often
parameterised via an average inter-chain exchange constant $J'$, are
extremely weak, as suggested by magnetic susceptibility measurements
\cite{Mennenga-1984} which implied $|J'/J|<10^{-2}$. Later muon-spin
relaxation measurements showed the presence of a magnetic phase
transition to a regime of three dimensional long-range magnetic order
(LRO) for temperatures below a N\'{e}el temperature of
$T_{\mathrm{N}}=110$~mK \cite{Lancaster-2006}, leading to a revised estimate
of $|J'/J|=4.4\times10^{-3}$.  As a result of the small size of this
ratio, the signature of the phase transition is not visible in
specific heat measurements at low temperatures
\cite{Hammar-1999,Sengupta-2003,Jornet-Somoza-2010}.  A recent
electron spin resonance (ESR) study \cite{Validov-2014} suggests
further that the inter-chain exchange coupling in the $c$-direction may
be the most important in determining the low-energy behaviour of the
system, as it causes pathways in the $ac$-plane to form a frustrated
triangular spin lattice.

The weak inter-chain magnetic interactions in this compound make it a
suitable subject for a comparative study of the different structure
methods of calculating the exchange constants.  Previously,
Journet-Somoza {\em et al.}  \cite{Jornet-Somoza-2010} (hereafter JS)
used the DFA \cite{Whangbo-2003,Nova-2011} to show that small changes
in bond lengths between temperatures of 158~K and 2~K cause the
magnetic exchange along inter-chain pathways to become significant at
the lower temperature.  The crystal structure at 2~K is not expected
to change significantly as the temperature is lowered further, and so
they argue that the topology of magnetic exchange paths that they find
is valid below $T_{\mathrm{N}}=110$~mK.  Dos~Santos {\em et al.}
\cite{DosSantos-2016} (hereafter DS) have carried out DFA calculations
using the crystal structure measured at 100~K using the same exchange
functional, along with a calculation of the strongest coupling using
the PA. Their results are consistent with those of JA.  It is notable
that several of the exchange constants calculated in these studies are
relatively small, of the order of 10$^{-2}$ times smaller than the
leading order exchange, and it is therefore possible that the values
of these constants could be sensitive to systematic errors.

\begin{figure}
\begin{center}
\includegraphics[width=\columnwidth]{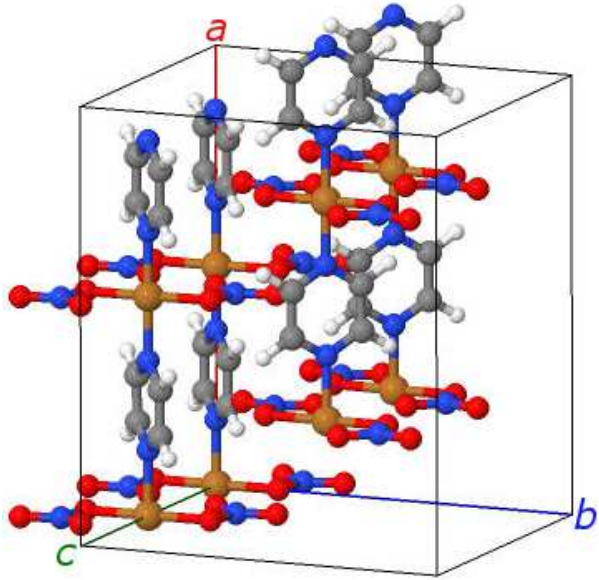}
\caption{Structure of the Cu(pyz)(NO$_3$)$_2$ unit cell at 
  $T=2$~K, as used in the PA calculation.  Gold indicates the Cu ions, red O,
  blue the N, grey C and white H.\cite{Jornet-Somoza-2010}.}
\label{fig:basicCell}
\end{center}
\end{figure}

The sensitivity of the exchange constants to numerical and systematic
error are examined here for Cu(pyz)(NO$_3$)$_2$ at 2~K within the framework
of a plane wave GGA+U \cite{Martin-2004} approach using both the DFA
and the PA. Below we discuss the implications of our results on the accuracy
of these approaches and the nature of the long-range magnetic order of
Cu(pyz)(NO$_3$)$_2$ at low temperature.

\section{Method}
To extract the Heisenberg coupling constants, 
we employ a collinear magnetic model where ordered spins are
constrained to adopt parallel or antiparallel configurations. 
Our approach
involves determining the energy differences between ordered spin
states that differ by a number of reversed spins.  An
underlying physical assumption, therefore, is that upon magnetically
ordering, the magnetic structure is constrained to be collinear. In
other words, the value of the magnetic exchange derived assumes that
nearest neighbor spins $\boldsymbol{S}$ at sites $i$ and $j$ obey $\boldsymbol{S}_{i}\cdot \boldsymbol{S}_{j}= \pm 1$. If this is not the
case, then the error in this assumption is absorbed into the value of
the exchange constant  that is derived. 
More specifically, we map the magnetic
centres of the system to an Ising Hamiltonian
\cite{Whangbo-2003,Nova-2011,Moreira-2006}
\begin{equation}
\hat{H}_{\rm Ising}=-2\sum_{i>j}I_{ij}S_{z\,i}S_{z\,j}.
\label{eqn:basicIsing}
\end{equation}
Here, $I_{ij}$ is the exchange constant parameterizing the interaction
between the magnetic centres (in this case the Cu$^{2+}$ ions) labeled
by $i, j$ and $\hat{S}_{z\,i}$ is the Ising spin operator for site
$i$.  The coupling constants $I_{ij}$ are calculated by relating the
energies of the ferromagnetic (FM) ordered state and the various
antiferromagnetic (AFM) ordered states found using the DFA or PA method.  We
convert an Ising coupling $I_{ij}$ for a given pathway to the desired Heisenberg
coupling $J_{ij}$ for that pathway using \cite{Datta-2015}
\begin{equation}
J_{ij}=\frac{I_{ij}}{N_{\rm mc}},
\end{equation}
where $N_{\rm mc}$ is the number of magnetic centres in the unit cell
(or supercell) used in the calculation.

\subsection{Dimer Fragment Approach}
\begin{figure}
\begin{center}
\includegraphics[width=6.3cm]{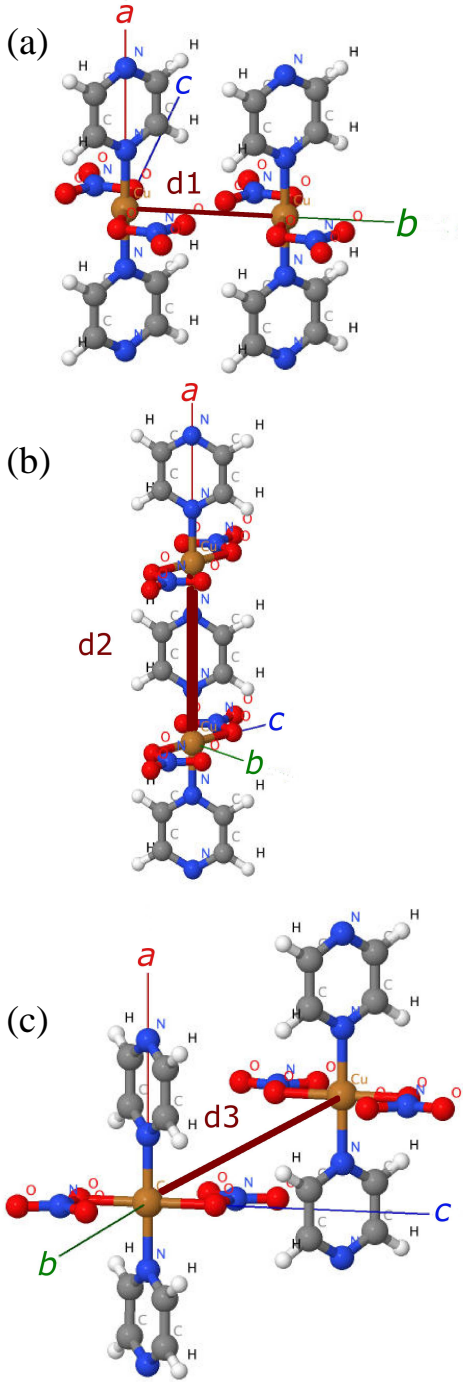}
\caption{Exchange pathways used in
  Ref.~\onlinecite{Jornet-Somoza-2010}.  (a) d1 (directed along the
  $b$-axis), (b) d2 (along the $a$-axis), (c) d3 (along
  the $c$-axis).
\label{fig:dimerModels}}
\end{center}
\end{figure}

In the DFA \cite{Whangbo-2003,Nova-2011,Moreira-2006}, the system is
divided into pairs of magnetic centres corresponding to magnetic
exchange pathways.  In many cases \cite{Moreira-2006} the value of the
exchange constant linking those centres can be described accurately
using only the two centres and their corresponding ligands. It is then
reasonable to assume that the value of the exchange constant may be
calculated using a model consisting of an isolated system that
includes only those components.  We select isolated dimer fragments
of the system that correspond to the three largest exchange pathways,
illustrated in Fig.~\ref{fig:dimerModels}, and obtain the value of the
exchange coupling, $J$, along a particular pathway by relating the FM
and AFM energies of the corresponding Ising dimer system via
\begin{equation}
J=\frac{E_{\rm AFM}-E_{\rm FM}}{2},
\label{eqn:dimerJ}
\end{equation}
where $E_{\rm AFM}$ is the energy of the dimer in an antiferromagnetic
spin configuration and $E_{\rm FM}$ is the energy of the ferromagnetic
configuration.

We use the labeling conventions of
Ref.~\onlinecite{Jornet-Somoza-2010}, where $J_{\mathrm{d}1}$ is the exchange
coupling along the $b$-direction, $J_{\mathrm{d}2}$ the exchange along the
$a$-direction and $J_{\mathrm{d}3}$ the exchange along the $c$-direction.  The
parameter $J_{\mathrm{d}2}$ is therefore the value of the superexchange
coupling between magnetic sites along the Cu-pyz-Cu chains and
determines the energy scale of the 1D behaviour of the system at
temperatures $T_N\le T\lesssim J$ \cite{SachdevBook}. The couplings
$J_{\mathrm{d}1}$ and $J_{\mathrm{d}3}$ contribute to the 3D
magnetic ordering behaviour of the system on cooling towards $T_{\mathrm{N}}=110$~mK.
The relative strengths of the three couplings will determine the
properties of the low temperature LRO. The structure they describe is
that of a stacked triangular lattice \cite{Collins-1997} and so
effects related to spin
frustration in the $ac$ plane may arise if $J_{\mathrm{d}3}/J_{\mathrm{d}2}$ is
sufficiently large.

\subsection{Periodic Approach}\label{sec:directMethod}
\begin{figure}
{\centering
\includegraphics[width=\columnwidth]{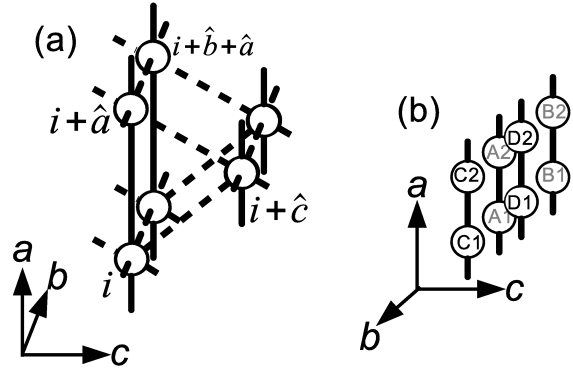}
\caption{(a) The Ising model mapping of the chain is illustrated
  showing the magnetic centres, superexchange pathways (full lines)
  and direct exchange pathways (broken lines).  Vectors $\hat{a}$,
  $\hat{b}$ and $\hat{c}$ indicate hopping in that direction by a
  single magnetic site.  (b) Schematic of Fig.~\ref{fig:basicCell}
  showing only the Cu$^{2+}$ ions, the superexchange pathways along
  the chains and the labeling convention used for spin
  configurations.}
\label{fig:IsingMap}}
\end{figure}

In the PA \cite{Whangbo-2003,Moreira-2006} we map the magnetic
structure of the compound to the model depicted in
Fig.~\ref{fig:IsingMap}(a) where $\hat{a}$, $\hat{b}$ and $\hat{c}$
give a shift by one site in the directions {\it a}, {\it b} and {\it c}
respectively. Calculations are performed for a periodic unit cell
containing eight Cu ions [see Figs.~\ref{fig:basicCell} and
  \ref{fig:IsingMap}(b)], which is the smallest number of Cu ions
needed to realize enough spin configurations to calculate
$J_{\mathrm{d}1},\,J_{\mathrm{d}2}$ and $J_{\mathrm{d}3}$.

Comparing Eq.~(\ref{eqn:basicIsing}) and Fig.~\ref{fig:IsingMap}(a),
we write the Ising Hamiltonian of the system as
\begin{equation}
\hat{H}= \hat{H}_{d1}+\hat{H}_{d2}+\hat{H}_{d3},
\end{equation}
where 
\begin{eqnarray}
\hat{H}_{d1}&=&-2I_{d1}\sum_{i=1}^n\hat{S}_{z\,i}\hat{S}_{z\,i+\hat{b}}, \nonumber\\
\hat{H}_{d2}&=&-2I_{d2}\sum_{i=1}^n\hat{S}_{z\,i}\hat{S}_{z\, i+\hat{a}}, \nonumber\\
\hat{H}_{d3}&=&-2I_{d3}\sum_{i=1}^n\left(\hat{S}_{z\,i}\hat{S}_{z\, i+\hat{c}}
+\hat{S}_{z\,i+\hat{c}}\hat{S}_{z\,i+\hat{a}}\right).
\end{eqnarray}
Here, $n$ is the total number of lattice sites.  Note that $H_{d3}$
explicitly defines a triangular exchange topology, as can be seen from
Fig.~\ref{fig:IsingMap}(a).

\begin{table}[t]
\centering
\begin{tabular}{ccc}
\hline\hline
State & Label & Energy\\
\hline
00000000 & FM & $E_{\rm FM}=-4J_{\mathrm{d}1}-4J_{\mathrm{d}2}-8J_{\mathrm{d}3}$\\
01101001 & AFM1 & $E_{\rm AFM1}=4J_{\mathrm{d}1}+4J_{\mathrm{d}2}$\\
01011010 & AFM2 & $E_{\rm AFM2}=4J_{\mathrm{d}1}+4J_{\mathrm{d}2}$\\
01100110 & AFM3 & $E_{\rm AFM3}=-4J_{\mathrm{d}1}+4J_{\mathrm{d}2}$\\
00111100 & AFM4 & $E_{\rm AFM4}=4J_{\mathrm{d}1}-4J_{\mathrm{d}2}+8J_{\mathrm{d}3}$\\
\hline\hline
\end{tabular}
\caption{Energies for spin configurations of the unit cell shown in
  Fig.~\ref{fig:basicCell}.  These states are degenerate with those
  resulting from the reversal of all spins.}
\label{tab:energies}
\end{table}
Using the labeling convention of Fig.~\ref{fig:basicCell} and
Fig.~\ref{fig:IsingMap}(b), we denote different ordered spin
configurations we have calculated as a list of 0s (spin down) and 1s
(spin up) in the order A$_1$A$_2$B$_1$B$_2$C$_1$C$_2$D$_1$D$_2$.
Table~\ref{tab:energies} lists trial ferromagnetic
(FM) and antiferromagnetic states (AFM1, AFM2, AFM3, and AFM4), and
how these are related to the exchange constants.  Note that AFM1 and
AFM2 are degenerate; we label their energy as $E_{\rm AFMG}$.  From
the expressions for the energy of the configurations,
the exchange constants \cite{Datta-2015} are obtained via
\begin{eqnarray}
J_{\mathrm{d}1}&=&\frac{E_{\rm AFMG}-E_{\rm AFM3}}{64},\nonumber\\
J_{\mathrm{d}2}&=&\frac{E_{\rm AFM3}-E_{\rm FM}}{64},\nonumber\\
J_{\mathrm{d}3}&=&\frac{E_{\rm AFM4}+E_{\rm AFM3}-E_{\rm FM}-E_{\rm AFMG}}{128}.
\label{eqn:directCouplings}
\end{eqnarray}

\subsection{Numerical Details}

\begin{table*}
\centering
\begin{tabular}{cccccc}
\hline\hline   \multirow{2}{*}{}  &\multirow{2}{*}{PA }  &
\multicolumn{3}{c}{DFA}\\\cline{3-5} & & d1 & d2 &d3 \\
\hline
{\em a} (\AA) & 13.383 & 21.440 & 28.631 & 25.286\\
{\em b} (\AA) & 10.211 & 21.788 & 16.683 & 16.683\\
{\em c} (\AA) & 11.600 & 19.397 & 19.394 & 27.194\\
\hline\hline
\end{tabular}
\caption{Lattice parameters and convergence criteria used for the
  calculations.}
\label{tab:settings}
\end{table*}

All calculations were carried out using the CASTEP electronic
structure package\cite{CASTEP} with accurate \cite{Lejaeghereaad-2016}
`on-the-fly' ultrasoft, PBE \cite{Perdew-1996} pseudo-potentials.  The
cell sizes used are summarised in Table~\ref{tab:settings}. The DFA
requires that the dimers used in the calculation are isolated. This
was simulated in a pseudo-periodic approach by choosing cell
dimensions so that each periodic image of the fragment is separated by
sufficient vacuum to be isolated, as discussed in the Supplemental
Information\cite{si}.

Recent work on the comparison of different implementations of DFT
\cite{Lejaeghereaad-2016} discuss the quantity $\Delta$, which is a
measure of the difference in converged quantities, particularly
energy, between different DFT implementations. These indicate that the
best $\Delta$ values across a range of DFT implementations are of the
order of 0.5~meV/atom therefore when examining small total energies
there are implementation differences which indicate a limit to the
absolute accuracy. However energy differences within a particular
implementation are more accurate as some error cancellation occurs.

To that end, total energy differences are converged with respect to
basis set size (Monkhorst-Pack (MP) grids\cite{Monkhorst-1976}, plane
wave cutoff and appropriate spatial discretisations) to a higher
tolerance than usual, being accurate to better than 0.1~meV/cell. This
is the value that limits the accuracy on energy differences and hence
coupling constants. The self-consistent eigenvalues are converged to
$10^{-12}$~eV to ensure that self-consistency in the calculation does
not impose additional numerical error. Details are given in the
Supplemental Information.

\begin{figure}
\begin{center}
\includegraphics[width=6.4cm]{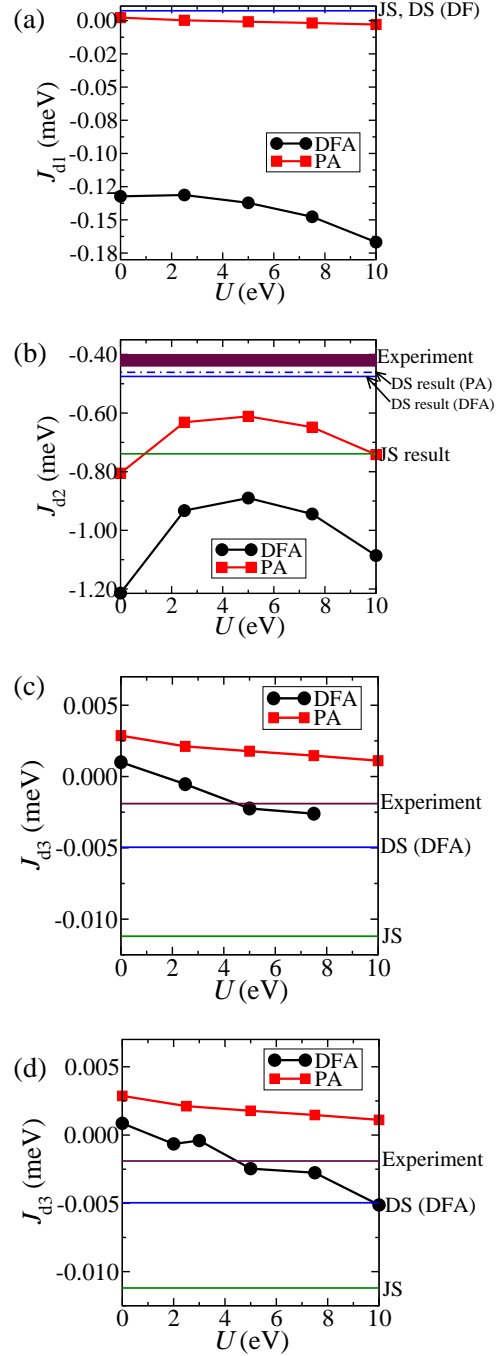}
\caption{Calculated exchange constants compared with other
  computational and experiment data for (a) $J_{\rm d1}$; (b) $J_{\rm
    d2}$; (c) $J_{\rm d3}$; (d) $J_{\rm d3}$ with a larger unit cell
  size. Experimental results in (b) are from magnetic susceptibility
  measurements \cite{Hammar-1999} and in (c) are ESR
  \cite{Validov-2014}.
\label{fig:Jvals}}
\end{center}
\end{figure}

\section{Results}
\subsection{Exchange Constants}
The differences we find in the calculated energies for the different spin structures are very small,
reflecting the weakness of the inter-chain exchange constants compared
to the dominant intra-chain exchange.  Figure~\ref{fig:Jvals} shows our
results for the exchange couplings compared to results from earlier
work \cite{Jornet-Somoza-2010,DosSantos-2016} and experimental
estimates \cite{Hammar-1999,Validov-2014}.  

First we examine the
largest exchange coupling $J_{\rm d2}$ [Fig.~\ref{fig:Jvals}~(b)],
which corresponds to the exchange along the Cu-pyz-Cu chains.  We
note that both our DFA and PA calculations both predict AFM
coupling and are comparable with both previous calculations and the
values derived from magnetic susceptibility
measurements.\cite{Hammar-1999} 

Although no single calculation
reproduces the experimental value, (i) the DS results, calculated
using both PA and DFA, provide the best agreement, followed by (ii)
our PA result evaluated with $U=5.0$~eV, (iii) the JS result and (iv)
our DFA calculation evaluated at $U=5.0$~eV.  For $0\le U < 5.0$~eV,
our PA and DFA values approach the experimental value as $U$ is
increased.  Above $U=5.0$~eV, we see that $J_{\rm d2}$ decreases,
departing from the experimental value as $U$ is further increased, and
that at $U=10.0$ eV the PA result is close to the JS value.  The best
agreement with experiment from our calculations is obtained at $U=5.0$
eV for both methods, with the PA giving the closer agreement.  Since
$U=5.0$~eV gives the best match of $J_{\rm d2}$ to the experimentally
measured values, we empirically fix this value to compare the
calculated exchange constants at $U=5$~eV to the results of previous
calculations in Table~\ref{table:Comparison}. 

We may summarize that the calculated values of the principal exchange
constant $J_{\mathrm{d2}}$ lie above the predicted limit of the energy
resolution of our well-converged calculations ($\approx 0.1$~meV/cell)
and the variation between our results and the previous ones lie within
the expected variation for different implementations ($\approx
0.5$~meV/atom).  It is also notable that the difference between our PA
and DFA results is also of this order. This is discussed further in
Section~\ref{discussion}

The most notable property of the other coupling constants
($J_{\mathrm{d1}}$ and $J_{\mathrm{d3}}$) is that
they are found to be very small compared to $J_{\mathrm{d2}}$. 
No experimental results
are available for $J_{\rm d1}$ but both JS
and DS predict FM coupling [Fig~\ref{fig:Jvals}(a)]. In contrast, our
DFA results indicates AFM coupling, while our PA calculation gives a coupling very
close to zero, with $J_{\rm d1}$ changing sign between
$U=2.5$~eV and $U=5.0$~eV,  going from FM coupling to AFM
coupling. 
The results for $J_{\rm d3}$ [Fig.~\ref{fig:Jvals}(c)] show that both
our PA and DFA exchange constants decrease with increasing $U$.  Our
PA data indicate a FM coupling while in the DFA a small FM coupling
that evolves into a small AFM coupling which, at $U=5.0$~eV, is close
the ESR-derived estimate\cite{Validov-2014}.  
[We note in passing that the  $U=10.0$ eV  result  for the  DFA
calculation  ($J_{\rm d3}=0.024$  meV)  is omitted  from Fig.~\ref{fig:Jvals}(c),
since its behaviour is a consequence  of the large
$U$ value giving the system a different ground state,
and cannot then be compared  with the
other  results.
  Increasing  the  vacuum  spacing  by
expanding  the cell  dimensions  in all  directions  by 2.5~\AA\  [Fig.
  \ref{fig:Jvals} (d)]  causes the $U=10.0$~eV value to  become an AFM
exchange constant.
The resulting similarity between the values in Fig.~\ref{fig:Jvals}(c)
and  Fig.~\ref{fig:Jvals}(d) justifies the use of the smaller cell.]

To summarize the results from the subdominant exchange couplings, we note that 
the absolute magnitude predicted for these constants is
small compared to the 0.1~meV/cell resolution limit we predict for the
calculations, leading us to doubt that they are meaningful. 
This is discussed in more detail below (see Section~\ref{discussion}), but immediately suggests that any attempt to derive even the qualitative behaviour of the
system will fail owing to the changes in sign of these quantities. 

\begin{table*}
\centering
\begin{tabular}{ccccc}
\hline\hline
\small Exchange &  \multirow{2}{*}{\small JS \cite{Jornet-Somoza-2010}}   &  \multirow{2}{*}{\small DS \cite{DosSantos-2016}}&
 {DFA} & PA\\ 
\cline{4-5}
\small pathway &   &  &\small $U=5.0$ eV   &\small $U=5.0$ eV  \\
\hline
\small $J_{\mathrm{d}1}$ ($\times10^{-4}$ eV) & \small  $0.07$ & \small $0.07$ &  \small $-0.14$& \small $-0.02$\\
\small $J_{\mathrm{d}2}$ ($\times10^{-4}$eV) & \small $-7.39$& \small $-4.61$ (PA), $-4.75$ (DFA) &\small $-8.90$& \small $-6.61$\\
\small $J_{\mathrm{d}3}$ ($\times10^{-4}$eV)  & \small  $-0.11$ & \small $-0.02$ &  \small $-0.04 $& 
\small $0.14$ \\
\hline\hline
\end{tabular}
\caption{Comparison of our results with those of
  Refs.~\onlinecite{Jornet-Somoza-2010} and
  \onlinecite{DosSantos-2016}.}
\label{table:Comparison}
\end{table*}

\subsection{Energy levels and band structures}
\begin{figure}
{\centering
  \includegraphics[width=\columnwidth]{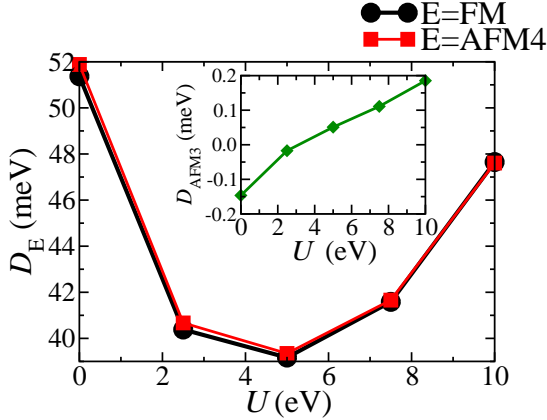}
\caption{Plots of the relative energies per supercell $D_{\rm
    E}=E_{\rm E}-E_{\rm AFMG}$.}
\label{fig:energies}}
\end{figure}
An immediate consequence of the small differences in the energies of 
the low-lying
 magnetic states is that it is difficult to reliably determine
the predicted ground state magnetic spin structure.
Fig.~\ref{fig:energies} shows the separation of energies $D_{\rm E}$
(where ${\rm E}$ labels the states ${\rm AFM3,\, AFM4,\, FM}$)
relative to the energy of the low-lying AFMG state.  We see that both
$D_{\rm AFM4}$ and $D_{\rm FM}$ follow roughly parabolic behaviour as
$U$ increases, with a minimum in energy difference of $D_{\rm
  AFM4,FM}\approx 0.039$~meV occurring at $U=5.0$~eV. These two
quantities are very similar in value at $U=0$ and become more so as
$U$ increases.  In comparison $D_{\rm AFM3}$ is found to be small
(less than $\pm 0.2$~meV) and is negative until a point
$U_{\mathrm{c}}$ between $U=2.5$~eV and $U=5.0$~eV, where it becomes
positive.  This implies that below $U_{\mathrm{c}}$ the magnetic
ground state structure of that system is predicted to be AFM3; while
it is predicted to be the
AFMG state above $U_{\mathrm{c}}$.

\begin{figure}
\begin{center}
\includegraphics[width=6cm]{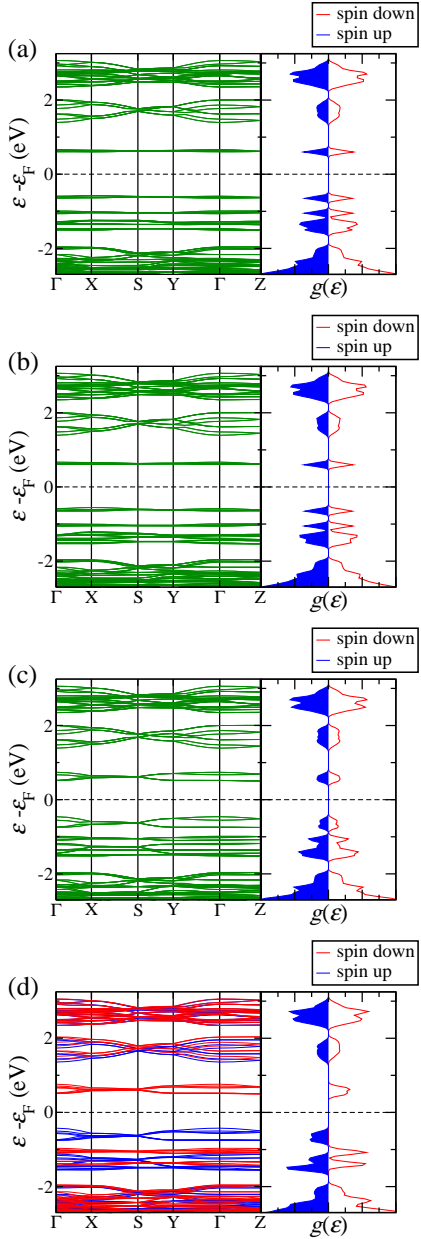}
\caption{Band structure  and density  of states  $g(\epsilon)$ for  (a)
  AFMG, (b) AFM3, (c) AFM4 and (d) FM magnetic states.
\label{fig:bandstruct}}
\end{center}
\end{figure}

Details of the electronic structure of these magnetic systems is shown
in the band structures at $U=5.0$~eV, displayed in
Fig.~\ref{fig:bandstruct}.  Qualitatively, the bands closest to the
Fermi energy in AFM3 [Fig.~\ref{fig:bandstruct} (b)] and AFMG
[Fig.~\ref{fig:bandstruct}(a)] have similar band structures and
density of states $g(\varepsilon)$, but with some small additional
splitting of degeneracies visible in the AFM3 bands. Apart from a
standard splitting in spin channel energies, the FM band structure
[Fig.~\ref{fig:bandstruct}(d)] shows that the states closest to the
Fermi energy are qualitatively very similar to AFM4
[Fig.~\ref{fig:bandstruct} (c)].  (In the latter case the density of
states plots are noticeably different, since the spins in the FM case
are only oriented in one direction.)  The band structures of the
AFM3/AFMG and FM/AFM4 groups are qualitatively different from each
other, which is consistent with the small energy separation between
the AFM3 and AFMG states on the one hand, and the FM and AFM4 states
on the other, as well as the large energy separation between the AFMG
and the AFM4 and FM states.

\section{Discussion}\label{discussion}
Although DFT+U yields results that are able to describe a range of
magnetic structures in this system, all calculations make use of
approximations that will give rise to error. From
Fig.~\ref{fig:energies} we see that this is particularly significant
for structures such as AFMG and AFM3 (or AFM4 and FM), which are very close in energy.
It is the energy differences in Eqs.~(\ref{eqn:directCouplings}) that determine
the $J$-couplings and hence it is the errors in these energy
differences that
determine the reliability of the results.  The magnitudes
of the energy differences between the FM and AFM4 states and the AFM3
and AFMG states are small (predicted to be fractions of 
1~meV/atom). When one compares these
to absolute energy convergence with respect to basis set (k-points,
energy cutoff and spacial grids), predicted to be of order 0.1~meV/cell, errors are both inevitable and
likely to be large relative to the calculated values of the small exchange
constants. Due to these considerations we have no grounds to expect
calculations using different implementations of the calculation or
different techniques to agree on the values
of these exchange constants.

An additional source of approximation in the DFA is that arising from
the truncation of the full crystal structure down to a pair of
magnetic centres.  This neglects the contributions to the exchange
constant that might arise from neighbouring magnetic centres.
Furthermore, the confinement of electrons within a smaller subsystem
of the chemical structure will tend to increase their kinetic energy
relative to that of electrons in the full structure, much as the
energy of an electron confined in a box is larger if the dimensions of
the box are made smaller.  These effects will only be significant if
they are of similar order to the magnitude of the exchange constant.
For the dominant exchange pathway this will not usually be the case,
which is why this approach can produce qualitatively accurate results
for these couplings that are comparable with the results of the PA
method [as seen in Fig.  \ref{fig:Jvals}(b) for $J_{\rm d2}$], and why
the trimer cluster calculation of JS \cite{Jornet-Somoza-2010} did not
produce a qualitatively different result from their dimer calculation.

It is clear that calculations of subdominant Heisenberg exchange
constants of the order of 0.01~meV calculated using a single method
and/or exchange-correlation functional should not be taken at face
value, as the calculations are not converged enough with respect to
basis set. A reliable conclusion that can be drawn is merely that
these exchange constants are small compared to the energy resolution
of the calculation method.

\section{Conclusion}
We have calculated the three largest magnetic exchange constants in
Cu(pyz)(NO$_3$)$_2$ using well-converged, plane wave density
functional methods, augmented by a Hubbard-U approach, using two
different structural models.  The results of both are qualitatively
consistent with each other and experiment for the dominant nearest
neighbour exchange constant $J_{\mathrm{d}2}$. However this does not
hold true for the smaller $J_{\mathrm{d}1}$ and $J_{\mathrm{d}3}$
exchange constants, for which different calculational approaches and
implementations of DFT may give qualitatively different results. This
is because the difference in energy between several magnetic states of
this system is small enough that it cannot be reliably resolved by
state-of-the-art DFT implementations. For the very small coupling
constants, we should not expect consistency between calculations that
use different functionals, for example as $U$ varies.  
The small exchange constants in
Cu(pyz)(NO$_{3}$)$_{2}$ determine the nature of the 3D LRO that occurs
below $T_{\mathrm{N}}=110$~mK.  Since the different techniques
discussed in this paper give different values of the inter-chain
exchange constants, they also imply different magnetic ground
states. To the extent that we are uncertain as to the value of these
exchange constants, we are uncertain as to the true LRO of the system
at low temperatures.

It is also worth noting that here that the use of hybrid density
functionals will not lead to any improvement in the accuracy of the
interchain couplings. (The results calculated in Refs~\onlinecite{DosSantos-2016}
and \onlinecite{Jornet-Somoza-2010} involved the use of such hybrid functionals.) We have demonstrated that the main source of
error in the calculation of coupling constants is numerical precision
and the numerical precision using hybrid functionals is significantly
worse than using density-based functionals. This is because in
calculating (and hence converging) total energies, the error is first
order on the wavefunction and second order on the density.
As an aside, it should be noted that nearly all hybrid functionals
have free parameters that are often empirically fitted.  These
unconstrained functionals sacrifice physical rigor for the flexibility
of such empirical fitting and have been shown to be becoming less
accurate with time \cite{Medvedev-2017}.

More generally, low dimensional molecule-based magnets are usually
characterized by one relatively large exchange constant $J$ that
determines the energy scale of the low-dimensional behaviour expected
to occur for $T_{\mathrm{N}} < T < J$, and several smaller ones that
will determine the ordering temperature $T_{\mathrm{N}}$ and the
magnetic ground state of the system. The smaller the values of the
subdominant exchange constants in a particular material, the more
successful a realization of a low-dimensional spin model. 
It is worth noting that the reliable experimental determination of the
small, subdominant exchange constants in molecular magnets is
generally quite difficult. It is often not possible to extract them
uniquely from fits of the temperature dependence of magnetic
susceptibility, for example, and so their magnitude must be estimated
from combining measurements of the magnetic ordering temperature with
the results of modeling (e.g. from quantum Monte Carlo simulations)
\cite{Lancaster-2006}. In cases where there are several small
couplings, perhaps differing in sign, more sophisticated fitting of
neutron scattering results could yield the different couplings,
although such measurements generally require large single crystals and
the deuteration of the material, which has been shown to subtly alter
the magnetic properties \cite{maitland}.
 In DFT+U
and related theoretical approaches, the extreme sensitivity of the small exchange
constants to errors (that do not affect the value of the leading order
constant) is likely to be a general problem in reliably applying DFT
to these systems. It is likely, therefore, that the values of many of
the smallest exchange constants determined using DFT methods
(including {\it ab initio} XC functionals, semi-empirical hybrid
functionals and DFT+U) are little more than an artefact of the
implementation or convergence criteria, rather than the result of
controlled approximations.

\section{Acknowledgments}
This work was funded by EPSRC and the John Templeton Foundation as
part of the Durham Emergence Project.  We would like to thank the UK
Car-Parrinello collaboration for computer time.  Calculations were
performed using the ARCHER, N8 Polaris, and the Hamilton (Durham) HPC
facilities.  Research data from this paper will be made available via
Durham Research Online.

\end{document}